# Photocatalytic activity, optical and ferroelectric properties of $Bi_{0.8}Nd_{0.2}FeO_3$ nanoparticles synthesized by sol-gel and hydrothermal methods

*Hamed Maleki*[*]

Faculty of Physics, Shahid Bahonar University of Kerman, Kerman, Iran

e-mail address: hamed.maleki@uk.ac.ir

**Keywords:** Multiferroics; photocatalytic activity; bismuth ferrite; sol-gel; Neodymium doping; hydrothermal.

**Abstract:** In this study, the effects of synthesis method and dopant Neodymium ion on the ferroelectric properties and photocatalytic activity of bismuth ferrite were studied. $BiFeO_3$ (BFO) and $Bi_{0.8}Nd_{0.2}FeO_3$ (BNFO) nanoparticles were prepared through a facile sol-gel combustion (SG) and hydrothermal (HT) methods. The as-prepared products were characterized by X-ray powder diffraction (XRD), Furrier transform infrared spectroscopy (FTIR) and transmission electron microscope (TEM) images. Both nanophotocatalysts have similar crystal structures, but the SG products have semi-spherical morphology. On the other hand, HT samples have rod-like morphology. TEM results indicated that the morphology of products was not affected by the doping process. The thermal, optical and magnetic properties of nanoparticles were investigated by thermogravitometry and differential thermal analysis (TG/DTA), UV-vis spectroscopy, and vibrating sample magnetometer (VSM). The ferroelectric properties of BNFO nanoparticles were improved compared to the undoped bismuth ferrite. The photocatalytic activity of as-synthesized nanoparticles was also evaluated by the degradation of methyl orange (MO) under visible light irradiation. The photocatalytic activity of nanoparticles prepared via sol-gel method exhibited a higher photocatalytic activity compared to powders obtained by hydrothermal method. Also substitution of Nd into the BFO structure increased the photocatalytic activity of products.

# 1. Introduction

Multiferroic materials have simultaneously the properties of ferroelectricity, ferromagnetism and also in some cases ferroelasticity [1,2]. These materials have attracted attention because of their potential applications in data storage, spintronic devices and sensors and photovoltaics and so on [3–8]. Bismuth ferrite is the only known multiferroics that has rhombohedrally-distorted perovskite structure with the space group $R_3C$ [4] and shows both ferroelectricity and ferromagnetism at room temperature (RT) [9,10]. $BiFeO_3$ (BFO) has the ferroelectric order below the Curie temperate $T_C \sim 1103K$ and antiferromagnetic behavior below the Neel temperature $T_N \sim 643K$ [11,12].

In recent years, BFO has received new attention due to its narrow optical band-gap (2.19 eV) and excellent chemical stability, which allows the photocatalytic activity under visible light [3,13–18]. BFO shows different catalytic activities such as oxidation of organic compounds and degradation of pollutants [19–21]. In addition, BFO nanoparticles are magnetic semiconductor materials which could be separable in aqueous organic media [14]. However, there are some difficulties for different applications of BFO. Weak ferroelectricity, remanent polarization, high leakage current density, poor ferroelectric reliability and inhomogeneous weak magnetization are some of these challenges [22–24].

Many theoretical and experimental studies of bismuth ferrite have been investigated to expand the applications and solve these problems hindering practical usage of BFO. In order to overcome such problems and improve photocatalytic activity of BFO, many modification have been investigated which includes doping rare earth elements instead of bismuth into the BFO structure [15,25–34]. In addition, it was investigated that the substitution of rare earth (RE) elements into the bismuth ferrite, can alter the photocatalytic activity [35].

In recent years, there has been many reports on preparation methods for BFO nanoparticles with a focus on photocatalytic applications, which includes sol-gel route [36,37], co-precipitation [38] and hydrothermal reaction [39–41]. However, as the morphology of nanoparticles affects the physical properties of bismuth ferrite, we are interested to study the effect of synthesis process on the photocatalytic activity of pure and Nd-doped bismuth ferrite, as well as other physical properties of BFO and BNFO nanoparticles. Although several researches on the structural, optical and multiferroic properties of bismuth ferrite has been done, few studies have investigated on the photocatalytic activity of Nd-doped BFO and the influence of synthesis process [42,43]. The goal of this study is to quantify the effect of Nd dopant on photocatalytic degradation rates of MO. Furthermore, BFO and BNFO were characterized for the structure, morphology, energy band-gap. Moreover thermal, ferroelectric, leakage current density and magnetic properties of samples are investigated and compared between SG and HT as-synthesized products.

## 2. Experimental method

### 2.1 Sol-gel preparation of pure and Nd-doped BiFeO$_3$

In the sol-gel method, stoichiometric amount of Bi(NO$_3$)$_3$.5H$_2$O, Fe(NO$_3$)$_3$.9H$_2$O and Nd(NO$_3$)$_3$.6H$_2$O were dissolved in deionized water separately. Meantime, ethylene glycol (EG) and 2-methoxyethanol were mixed under stirring. Then, acetic acid was added to the solution dropwise (The pH of the mixed solution was adjusted to 1.5). This solution was mixed together under vigorous stirring for 30 min. Then metal nitrate solutions were mixed with fuel solution under constant stirring. The mixture was stirred constantly for 30 min and a dark red mixture appears. After stirring at RT for an hour, the temperature was increased to 70 ºC. After 3 hours heating and stirring, a clear brownish gel was obtained and following a few minutes with a temperature higher than 90 ºC, a yellow suspension is formed. The suspension was kept at RT for 10 hours and then it was put at 115 ºC for the water evaporation and fuel combustion. Finally the obtained powder was calcined at 650 ºC for 5 hours before investigating the characteristics.

### 2.2 Hydrothermal synthesis of BFO and BNFO

In the next part, BiFeO$_3$ and Bi$_{0.8}$Nd$_{0.2}$FeO$_3$ nanoceramics were synthesized by a hydrothermal process. Typically, bismuth (III) nitrate pentahydrate, ferric (III) nitrate nonahydrate, and neodymium nitrate hexahydrate (for BNFO) were dissolved in the minimum amount of deionized water as a specified stoichiometric ratio. The mixture wad dropped into potassium hydroxide (4M, 30 ml) under magnetic stirring. After stirring for 30 min, the mixture was placed in a teflon-lined steel autoclave of 50 ml for the hydrothermal reaction with a filling capacity of 80 % and performed at 200ºC for 12 hours in an oven and then cooled to RT naturally. The products are collected and washed several times with distilled water and ethanol and dried at 110 ºC for 2 h before further characterization.

### 2.3 Characterization

The structural properties of pure and Nd-doped BFO nanoparticles which are synthesized via both sol-gel and hydrothermal methods, were characterized by using XRD analysis (Philips powder diffractometer) with Cu-Kα radiation (λ=1.5406 Å) and Furrier transform inferared (FTIR) TENSOR27 spectrometer. Crystal sizes also were determined by the Scherrer method. Transmission electron microscope (TEM Leo-912-AB) was performed to study the morphology and size of products. The thermal behavior of the as-prepared samples are monitored by thermogravitometric and differential thermal analysis (TG/DTA NETZSCH- PC Luxx 409) with the heating rate of 10ºC/min up to 1000ºC. For the magnetic properties, the hysteresis loops are recorded up to 20 KG with the vibrating sample magnetometer (VSM- Lake Shore model 7410, SAIF) at RT. The optical properties of products were studied by UV-vis absorption spectra by using Lambda900 spectrophotometer. The polarization electric field P-E hysteresis loops of the prepared pellets were measured by the Sawyer- Tower circuit.

**2.4 Photocatalytic activity measurements**

The photocatalytic activity of the BFO and BNFO nanoparticles for decomposition of methyl orange (MO) was studied under irradiation of visible light source at the natural pH value. The reaction temperature was also kept at RT. The initial concentration of MO was 15 mgl$^{-1}$ with dispersing 0.1g BFO or BNFO in 200ml aqueous solution. Before irradiation, in order to reach an adsorption equilibrium of MO on products surface, the aqueous suspension was magnetically stirred for 75 min in the dark. Then the lamp was turned on and changes of MO concentration were measured by measuring the absorbance of the solution at 554 nm using a UV-vis spectrophotometer. C/C$_0$, where C was the concentration of MO at time t, and C$_0$ was the initial concentration of MO, was the photocatalytic degradation ratio of MO which has been investigated in this study.

# 3. Results and discussion
## 3.1 X-ray diffraction investigation

Figs. 1 (a) and (b) show the powder XRD patterns of BiFeO$_3$ (x=0) and Bi$_{0.8}$Nd$_{0.2}$FeO$_3$ (x=0.2) nanoparticles prepared by SG method. Analysis of patterns indicated that all products have a single perovskite phase with distorted rhombohedral structure with the space group R$_3$C. All samples reveal peaks that can be assigned to the standard card of BFO perovskite structure (JCPDS card No. 86-1518). By substituting 20% neodymium, a slight shifting of peaks towards lower angles occurred and a phase transformation from rhombohedral to tetragonal structure was observed (two major peaks at 30<2θ<33 merged into a one peak) and intensity of peaks were decreased. The width of the BFO peaks also increased with merging the nearby peaks. The size of products was calculated by using the Scherrer formula: $D = \frac{K\lambda}{\beta \cos\theta}$, where $K$ is the shape factor that normally measures to be about 0.89, λ stands for the wavelength of X-ray source, β is the width of the observed diffraction peak at its half intensity maximum, and θ is the Bragg angle of each peak. The obtained average nanocrystal sizes were 51 and 46 nm for BFO and BNFO respectively.

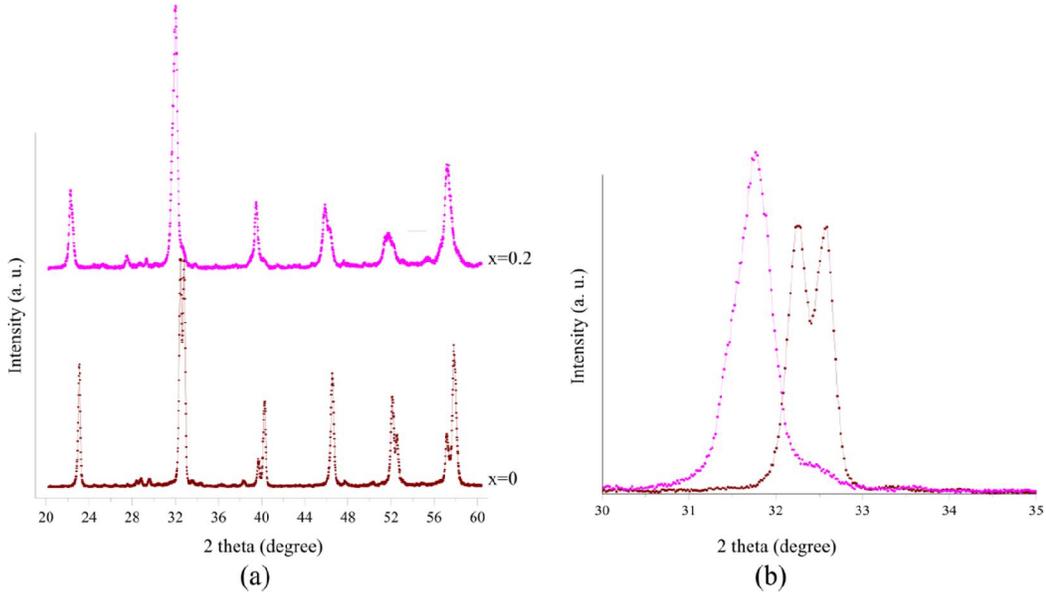

Fig. 1. (a) XRD patterns of $BiFeO_3$ and $Bi_{0.8}Nd_{0.2}FeO_3$ nanoparticles prepared by SG method. (b) Same graph in the range of $30<2\theta<33$

XRD patterns of as-prepared samples by HT are depicted in Figs. 2 (a) and (b). Analysis of diffraction patterns for the samples confirmed the perovskite single phase. All the major peaks of the patterns were possible to index in the rhombohedral phase ($R_3C$) for $BiFeO_3$. However for the case of BNFO, there is a shift in main peaks at $30<2\theta<33$ and furthermore, it has the tendency to merge in order to form one single widened peak (Fig. 2 (b)). One can relate this behavior to a smaller ionic radius in neodymium compared to bismuth. Moreover, comparison of Figs 1(b) and 2(b) indicates that width of major peacks in HT products are a little sharper compared to the as-synthesized SG samples.

Finally, the analysis of particle size, pointed out that in the presence of Nd, the average size of nanoparticles decreases (~40 nm for BFO and ~36 nm for BNFO). Neither the characteristic peaks of $Bi_2Fe_4O_9$ nor of $Bi_{25}FeO_{40}$ were found in any of BFO and BNFO patterns.

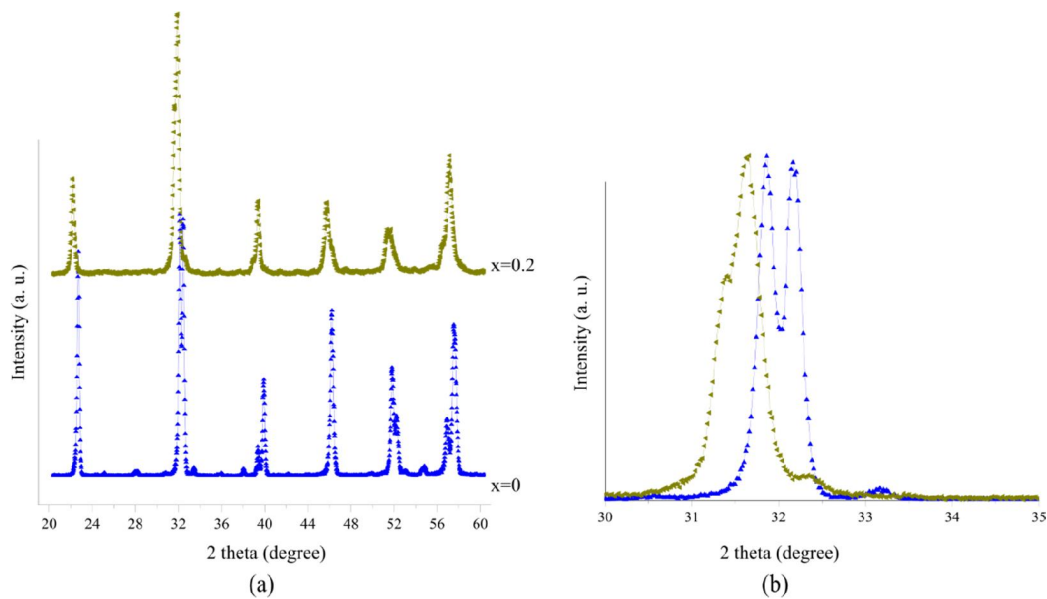

**Fig. 2.** (a) XRD patterns of pure and Nd-doped BiFeO$_3$ with x= 0.2 synthesized by HT route and (b) XRD patterns of pure BFO and BNFO in the range of 30<2θ<33.

### 3.2 Furrier transform infrared spectroscopy

In order to further confirm the crystallinity of as-prepared products, Fig. 3 shows the FT-IR spectra of BFO and BNFO nanoparticles in the range of 400-4000 cm$^{-1}$. In both SG and HT cases, analysis of curves and band range 400-600 cm$^{-1}$ is related to metal- oxygen bond and confirms the existence of perovskite structure for all samples. The vibration of Fe-O at ~450 cm$^{-1}$ and stretching vibration of O-Fe-O bonds at ~550 cm$^{-1}$ present in the octahedral FeO$_6$ group and in the framework can be observed below 600 cm$^{-1}$[44]. The broad band at 3200-3600 cm$^{-1}$ is due to antisymmetric and symmetric stretching of H$_2$O and O-H bond groups [45]. A strong band at around 1380 cm$^{-1}$ was due to the presence of trapped nitrates [46].

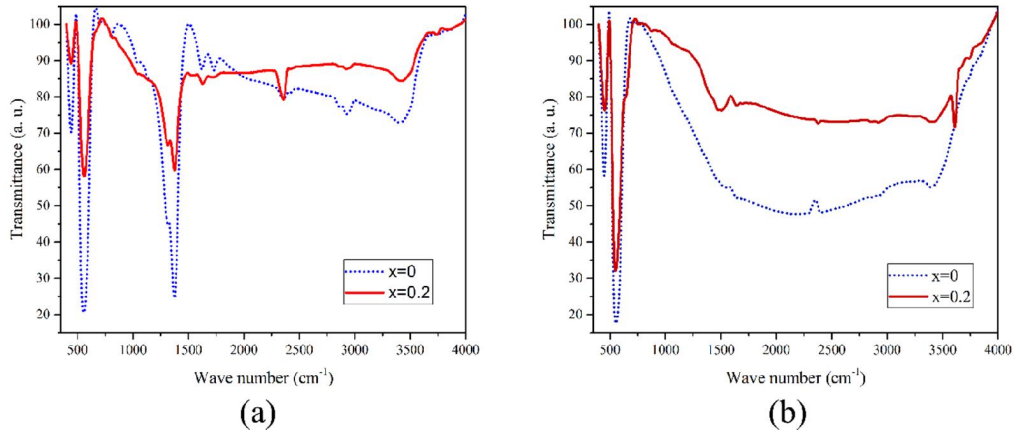

**Fig. 3.** (a) FTIR spectra of BFO and BNFO nanoparticles synthesized by SG and (b) products synthesized by HT.

### 3.3 Transmission electron microscope

Transmission electron microscope (TEM) is used for the observation of morphology and particle distribution and size of BFO and BNFO nanoparticles. As shown in Fig. 4, all samples consisted of nano-scale particles, however the morphology of powders depending on the synthesis method is very different. Although nanoparticles which were synthesized by SG are semi-spherical, products obtained via HT are nanorods. Figs. 4 (a) and (b) are BFO and BNFO nanoparticles that were prepared by SG method. Particles in both images are semi-spheroid and rectangular particles with some irregular shape particles which look denser. In contrast, products which were obtained from HT method have rod-like shape and look looser (Figs. 4 (c) and (d)). After incorporating the BiFeO$_3$ with Nd, it can be seen that there is no significant change in the morphology of both SG and HT products. The size of nanoparticles obtained from TEM is a little larger than that obtained from the Scherrer equation. According to analysis of images, the average size of SG as-prepared products were obtained 55 and 51 nm for BFO and BNFO nanoparticles and a thickness length ranging from 45-50 nm and 39-42 nm for HT as-prepared BFO and BNFO powders. The length of nanorods was also in a range of 80-300 nm. The results demonstrate a clear influence of the preparation method on the structural features of the prepared nanocomposites.

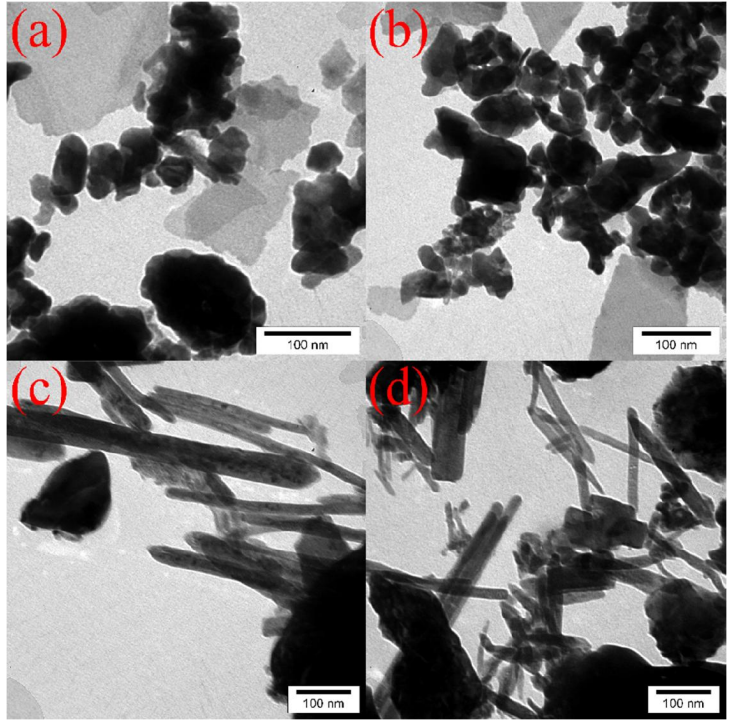

**Fig. 4.** (a) and (b) TEM images of BiFeO$_3$ and Bi$_{0.8}$Nd$_{0.2}$FeO$_3$ synthesized by SG method. (c) and (d) show BFO and BNFO nanopowders prepared via HT route.

### 3.4 Thermal behavior

The results of differential thermal analysis (DTA) for pure and 20%Nd-doped bismuth ferrite nanoparticles synthesized by the sol-gel method are shown in Figure 5 (a). The inset also indicates the thermogravimetric (TG) analysis of samples. In the case of BFO, in DTA curve for a temperature in the range of 180 °C, an exothermic peak can be seen with an instance of 0.2% reduction in weight. This peak can be attributed to hydrate thermal and the nitrate present in the course of evaporation as well as evaporating water on nanoparticle's surface. In the range of 400-600°C, one can observe a small exothermic peak as well as a weight reduction of 0.15%. This peak is an indication of the oxidation reaction between Fe$^{3+}$ and Bi$^{3+}$ and it is considered as an evidence for the crystal phase of BFO [47]. In the range of 820-840 °C, an endothermic peak appears which is due to electrical transmission temperature (T$_C$ ~ 830 °C). For Bi$_{0.8}$Nd$_{0.2}$FeO$_3$ in DTA curve, the hydrate and nitrate decomposition peaks were eliminated and 0.15% of the weight was declined. Also the phase transition from ferroelectric to Para-electric states occurred with exothermic peak in lower temperature (T$_C$ ~828 °C).

Inset in Fig. 5 (a) shows TG analysis of the BFO and BNFO nanoparticles. In the curves, up to 200°C the loss of physisorbed water is observed. The second part (200-450°C) corresponds to the loss of surface hydroxyl groups. Finally the weight loss is observed above 500°C, which is due to the decomposition of the nitrate spices [45].

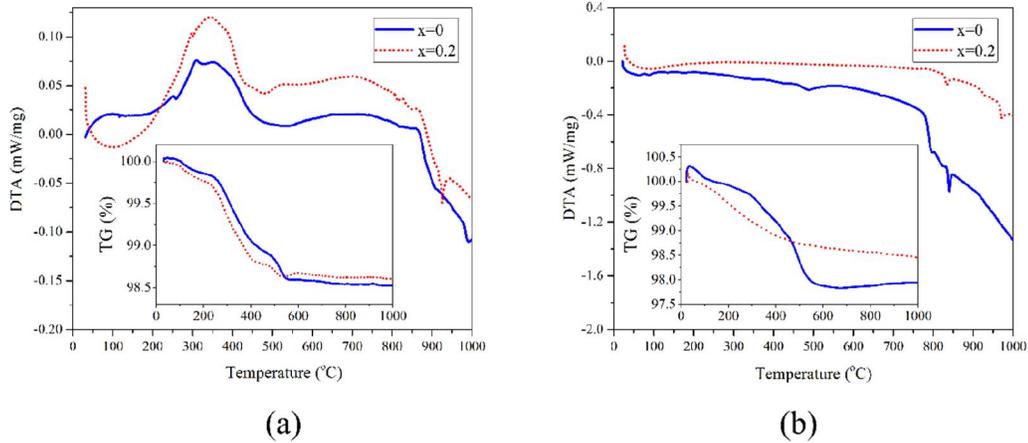

**Fig. 5.** (a) DTA and TG (as an inset) curves of BFO and BNFO nanoparticles synthesized by sol-gel method. (b) Same curves for as-prepared BFO and BNFO nanoparticles synthesized by hydrothermal method.

Same as Fig. 5 (a), in Fig.5 (b) differential thermal analysis curves of $BiFeO_3$ and $Bi_{0.8}Nd_{0.2}FeO_3$ nanocrystals are shown. These substances are synthesized by hydrothermal method. In DTA curves for a temperature at the range up to 200°C, an exothermic peak can be observed. This peak is due to available nitrate and evaporating water on the surface of nanoparticles. In the range of 400-600°C a small exothermic peak is seen. This peak indicates the oxidation reaction between $Fe^{3+}$ and $Bi^{3+}$. This can confirm the existence of crystal phase for BFO and 20%Nd-doped BFO. A phase transition at 833 °C for $BiFeO_3$ and 831 °C for $Bi_{0.8}Nd_{0.2}FeO_3$ can be seen in the measurements, which is related to the Curie temperature.

### 3.5 M-H hysteresis loops analysis of BFO and BNFO

The RT magnetic hysteresis loops for all samples are shown in Fig. 6. In contrast to the antiferromagnetism for bulk BFO, nanoparticle samples showed weak ferromagnetic behavior which is in agreement with other reports of BFO [10,48,49]. Table 1 shows the information of magnetic characterization tests for the as-prepared BFO and BNFO nanoparticles synthesized by SG and HT methods. According to literature, in bismuth ferrite the iron ions ($Fe^{3+}$) have a strong relationship with magnetization and these ions are responsible for magnetic properties of $BiFeO_3$. Surrounding each ($Fe^{3+}$) ion with a certain spin there are six other ions with nonparallel spins. These spins are not completely nonparallel however they are organized in a spiral manner with a period of 62nm, which leads to a magnetization value zero. Breaking the spiral organization of the spin is due to nanoparticles size reduction to less than 62 nm and the rise of uncompensated spins on the surface of nanocrystals (because of a rise of area relative to volume). This can be a justification for the increase in magnetic properties and the decline in ferromagnetic properties of bismuth ferrite [5,10]. When neodymium enters in the structure of bismuth ferrite, saturated magnetization ($M_s$) decreases. However, coercive force ($H_c$) increases. The remanent magnetization in HT-prepared samples is also increased, but for the SG-prepared samples, $M_r$ decreased.

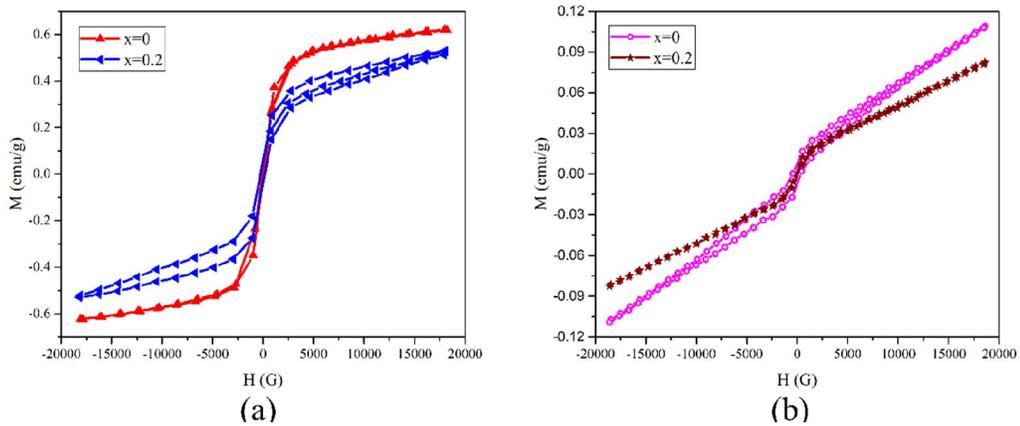

**Fig. 6.** (a) Magnetic hysteresis loops of BNFO nanoparticles with x=0,0.05,0.1,0.15,0.2 at two calcination temperatures of 550°C. (b) 650°C

By neodymium doping, the size of nanoparticles has declined. On the other hand, the morphology of nanoparticles are completely different according to synthesis methods. In general by decreasing the size of particles and domains, the energy for changing the magnetic moments increases, which is due to changes in mechanism of process. For the HT-prepared samples, loops are not saturated up to 20 KG. The saturated magnetization in SG-prepared samples is much higher compare to the other ones. On the other hand, the coercive force for samples obtained from HT method, is slightly higher than the SG prepared samples.

Table 1. Information from magnetic characterization for BFO and BNFO synthesized by sol gel and hydrothermal

| Sample | Sol gel | | | Hydrothermal | | |
|---|---|---|---|---|---|---|
| | $M_s$(emu/g) | $M_r$(emu/g) | $H_c$(G) | $M_s$(emu/g) | $M_r$(emu/g) | $H_c$(G) |
| $BiFeO_3$ | 0.63 | 0.0027 | 63.28 | 0.1093 | 0.0074 | 208.97 |
| $Bi_{0.8}Nd_{0.2}FeO_3$ | 0.54 | 0.0068 | 167.90 | 0.0096 | 0.0037 | 332.69 |

### 3.6 Ferroelectric properties of $BiFeO_3$ and $Bi_{0.8}Nd_{0.2}FeO_3$ nanoceramics

In order to study the ferroelectric properties of products, here, the as-synthesized BFO and BNFO nanoparticles were pressed and were coated with a thin layer (~30 nm) of silver as electrodes by using DC sputtering. Fig. 7 presents the RT polarization-electric field (P-E) hysteresis loops of BFO and BNFO nanoparticles synthesis by SG and HT methods under applied electric field up to 25 kV/cm. It can be seen that for all products loops are unsaturated. By adding Nd, into the structure of BFO, the saturated polarization is increased. However in both synthesis method results, remanent polarization and coercive field are decreased and reduced the leakage current [50,51]. The reduction of leakage current by doping Nd, is due to a reduction of oxygen vacancies [52,53]. When Nd is doped to BFO, the P-E loop is improved showing an elongated loop compared to the BFO samples.

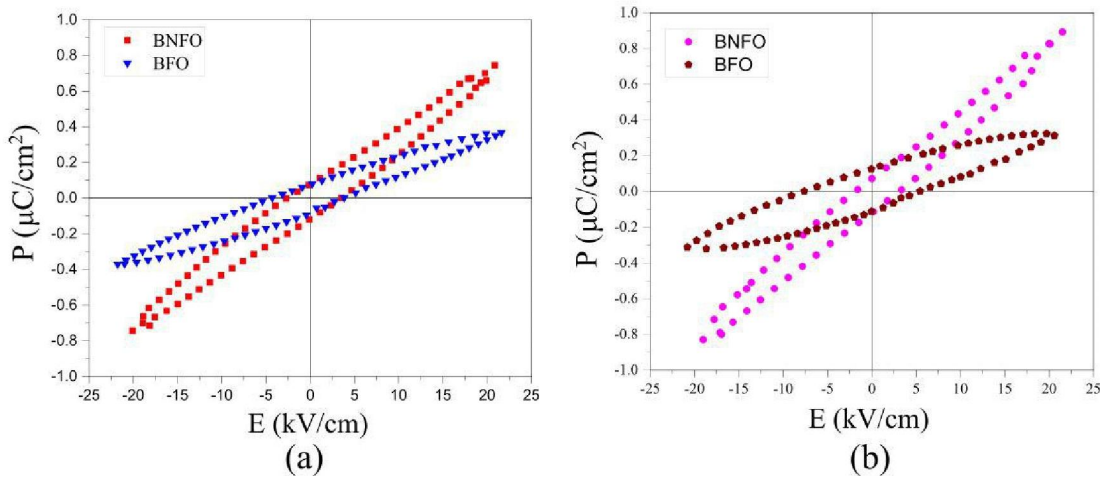

**Fig. 7.** P-E hysteresis loops of BFO and BNFO nanoparticles synthesized by (a) sol-gel and (b) hydrothermal.

In order to study the influence of Nd on the ferroelectric behavior of all products, leakage current density- electric field (J-E) curves of BiFeO$_3$ and Bi$_{0.8}$Nd$_{0.2}$FeO$_3$ nanoparticles synthesized by SG and HT at RT is plotted in Fig. 8. BFO (BNFO) nanoparticles synthesized via SG show high leakage current compared to the BFO nanorods which were synthesized by HT. Although the J is decreased with the dopant Nd in the nanoparticles and remained much lower than that of BFO, the synthesis method also affects the leakage current and in the case of BNFO, the products synthesized by HT route shows the bigger leakage current density.

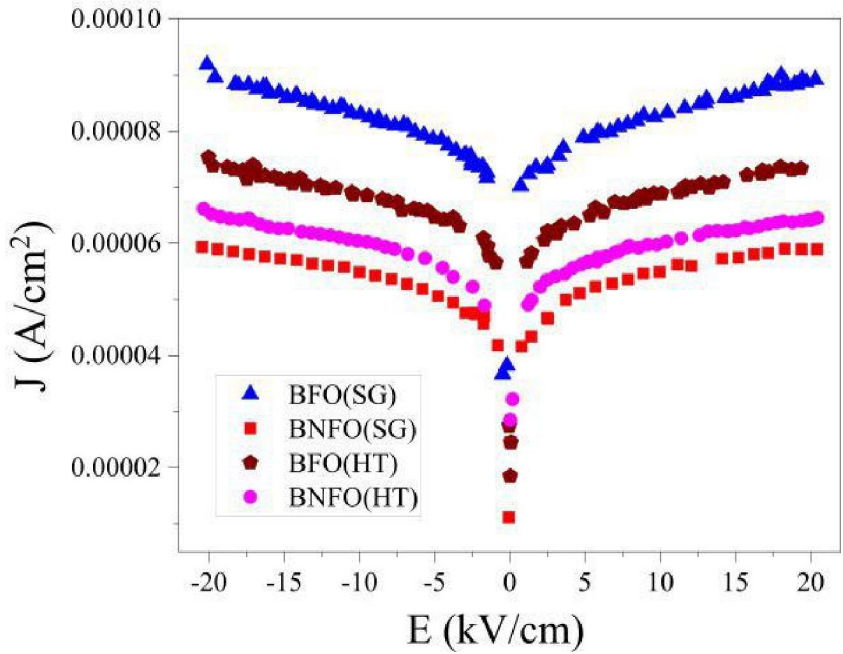

**Fig. 8.** RT current density- applied electric field (J-E) BFO and BNFO nanoparticles synthesized by SG and HT.

### 3.7 Uv-vis spectroscopy analysis

The optical band-gap of BiFeO$_3$ and Bi$_{0.8}$Nd$_{0.2}$FeO$_3$ nanoparticles have been calculated with the help of absorbance spectra. Fig. 9 shows the UV-vis absorption spectra of all products. By using Tauc's equation, the energy band-gap (E$_g$), absorption coefficient (α), is related by (for materials with direct band-gap): (αhν)$^2$=$K$(hν-E$_g$), where $K$ is a constant and hν is the photon energy. An inset in Fig.8 indicates the plot of (αhν)$^2$ vs hν for BiFeO$_3$ and Bi$_{0.8}$Nd$_{0.2}$FeO$_3$ nanoparticles synthesis by SG and HT methods. The extrapolated straight line fitted to the linear part of curves gives the value of E$_g$. The extracted values of E$_g$ for BFO (BNFO) nanoparticles synthesis by SG method is ~ 2.13 eV (2.08 eV) and 2.11 eV (2.04 eV) for products that obtained from HT reaction route. Slight decrease of $E_g$ value by Nd doping was observed which indicates the narrowing of the optical band-gap and enhanced photocatalytic activity and photovoltaic effects.

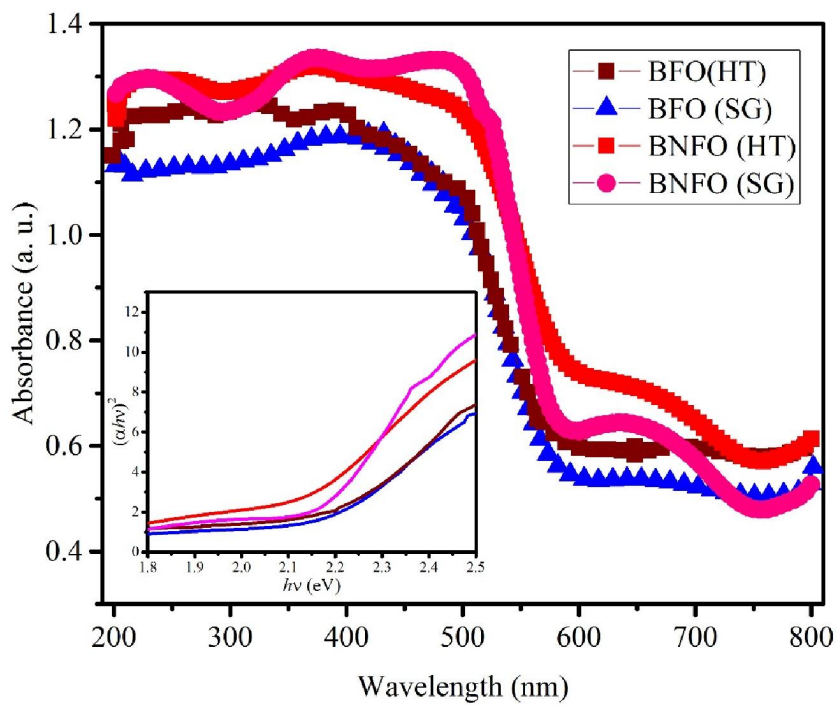

**Fig. 9.** UV-vis spectrum of $BiFeO_3$ and $Bi_{0.8}Nd_{0.2}FeO_3$ nanoceramics prepared by SG and HT methods. Inset shows $(\alpha h\nu)^2$ versus $h\nu$ plots of all samples.

### 3.8 Photocatalytic activity of BFO and BNFO

The photocatalytic activity of products were evaluated by degrading methyl orange (MO) in an aqueous solution under visible light irradiation. The reaction rate constant of samples are plotted in Fig. 10. All BFO and BNFO nanoparticles have visible light induced photocatalytic activity. The Nd-substitution has a significant effect on the photocatalytic activity of bismuth ferrite. The first-order rate constant is calculated by the equation $\ln(C_0/C) = \kappa t$, where $C_0$ and C are concentrations of MO in solution at the beginning of the tests and at time t. The reaction rate constant ($\kappa$) is the slope in the apparent of $\ln(C_0/C)$ vs. time. The HT-synthesized (SG-synthesized) BFO nanoparticles showed ~ 16% (23%) degradation of MO in 350 min. On the other hand, the addition of neodymium inside BFO nanoparticles, enhanced the photocatalytic activity of BFO nanopowders. The total removal of MO by HT-synthesized (SG-synthesized) BNFO is 61% (73%). This result indicates that the morphology of nanoparticles is an important factor that has effects on the photocatalytic activity of BFO or BNFO particles.

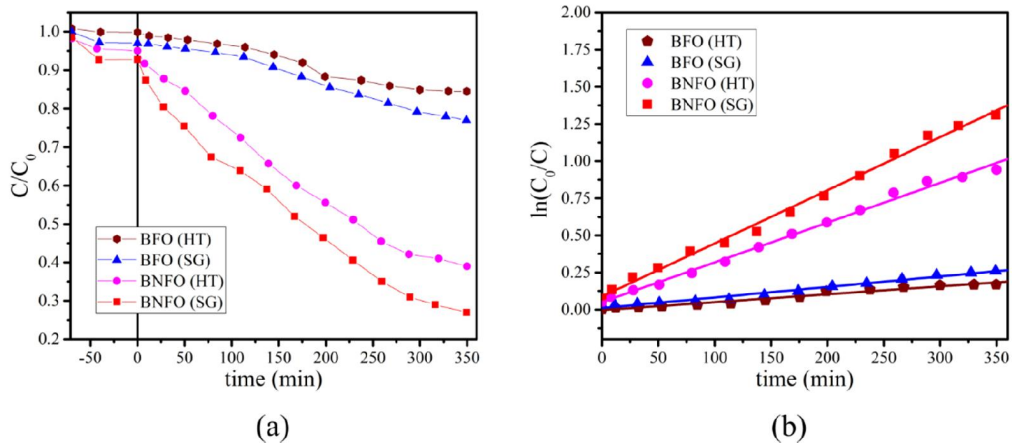

**Fig. 10.** (a) Photocatalytic degradation rates of BFO and BNFO nanoparticles prepared by SG and HT methods under visible light irradiation. (b) Represents the pseudo-first-order kinetics curves of the MO degradation.

## 4. Conclusions

In this paper the influence of synthesis method on the physical properties and photocatalytic activity of $BiFeO_3$ and $Bi_{0.8}Nd_{0.2}FeO_3$ nanoparticles was investigated by different characterization methods. In the BFO samples the main peaks change their shape and location after neodymium is involved and they have a tendency to merge and form a unit peak. The results from TG/DTA showed that by doping nanoparticles of bismuth ferrite the peak related to phase transition from ferroelectric to para electric took place in a lower temperature. The photocatalytic degradation of methylene orange (MO) under visible light irradiation was also implemented. Moreover the effect of neodymium doping into the bismuth ferrite, is studied. Results showed that the Nd-substitution, enhanced the ferroelectric characteristics and reduced the leakage current. In addition ferromagnetic properties were depended to the synthesis method which is related to the morphology of nanoparticles and by substitution Nd, the saturation magnetization was decreased.